# Roton pair density wave in a strong-coupling kagome superconductor

*Hui Chen[1,2,3,4#], Haitao Yang[1,2,3,4#], Bin Hu[1,2#], Zhen Zhao[1,2], Jie Yuan[1,2], Yuqing Xing[1,2], Guojian Qian[1,2], Zihao Huang[1,2], Geng Li[1,2,3], Yuhan Ye[1,2], Sheng Ma[1,2], Shunli Ni[1,2], Hua Zhang[1,2], Qiangwei Yin[5], Chunsheng Gong[5], Zhijun Tu[5], Hechang Lei[5], Hengxin Tan[6], Sen Zhou[7,2,3], Chengmin Shen[1,2], Xiaoli Dong[1,2], Binghai Yan[6], Ziqiang Wang[8*], and Hong-Jun Gao[1,2,3,4*]*

[1] *Beijing National Center for Condensed Matter Physics and Institute of Physics, Chinese Academy of Sciences, Beijing 100190, PR China*

[2] *School of Physical Sciences, University of Chinese Academy of Sciences, Beijing 100190, PR China*

[3] *CAS Center for Excellence in Topological Quantum Computation, University of Chinese Academy of Sciences, Beijing 100190, PR China*

[4] *Songshan Lake Materials Laboratory, Dongguan, Guangdong 523808, PR China*

[5] *Beijing Key Laboratory of Optoelectronic Functional Materials & Micro-Nano Devices, Department of Physics, Renmin University of China, Beijing 100872, PR China*

[6] *Department of Condensed Matter Physics, Weizmann Institute of Science, Rehovot, Israel*

[7] *CAS Key Laboratory of Theoretical Physics, Institute of Theoretical Physics, Chinese Academy of Sciences, Beijing 100190, China*

[8] *Department of Physics, Boston College, Chestnut Hill, MA, USA*

[#] These authors contributed equally to this work

[*] Correspondence to: wangzi@bc.edu

hjgao@iphy.ac.cn

**The transition-metal kagome lattice materials host frustrated, correlated, and topological quantum states of matter[1-9]. Recently, a new family of vanadium-based kagome metals $AV_3Sb_5$ (A=K, Rb, and Cs) with topological band structures has been discovered[10,11]. These layered compounds are nonmagnetic and undergo charge density wave transitions before developing superconductivity at low temperatures[11-19]. Here we report the observation of unconventional superconductivity and pair density wave (PDW) in $CsV_3Sb_5$ using scanning tunneling microscope/spectroscopy (STM/STS) and Josephson STS. We find that $CsV_3Sb_5$ exhibits a V-shaped pairing gap $\Delta$~0.5 meV and is a strong-coupling superconductor ($2\Delta/k_BT_c$~5) that coexists with $4a_0$ unidirectional and $2a_0\times2a_0$ charge order. Remarkably, we discover a 3Q PDW accompanied by bidirectional $4a_0/3$ spatial modulations of the superconducting gap, coherence peak and gap-depth in the tunneling conductance. We term this novel quantum state a roton-PDW associated with an underlying vortex-antivortex lattice that can account for the observed conductance modulations. Probing the electronic states in the vortex halo in an applied magnetic field, in strong-field that suppresses superconductivity, and in zero-field above $T_c$ reveals that the PDW is a primary state responsible for an emergent pseudogap and intertwined electronic order. Our findings show striking analogies and distinctions to the phenomenology of high-$T_c$ cuprate superconductors, and provide groundwork for understanding the microscopic origin of correlated electronic states and superconductivity in vanadium-based kagome metals.**

High-quality as-grown $CsV_3Sb_5$ crystals (Methods and Fig. S1) exhibit a stacking structure of Cs-Sb2-VSb1-Sb2-Cs layers with hexagonal symmetry (space group P 6/mmm) (Fig. 1**a**). In the VSb1 layer, the kagome net of vanadium is interwoven with a simple hexagonal net formed by the Sb1 atoms (Fig. 1**b**). The Cs and Sb2 layers form hexagon and honeycomb lattices, respectively (Fig. 1**b**). Samples characterized and the bulk electronic properties determined by temperature dependent magnetization, resistivity, and heat capacity (Methods). All measurements indicate the presence of an anomaly at $T \sim 94$ K (Fig. S2) associated with the charge density wave (CDW) transition[11,17]. At low temperatures, the magnetization, resistivity, and heat capacity measurements show a transition to the superconducting (SC) state at a critical superconducting temperature ($T_c$) of ~ 2.8 K. The somewhat higher $T_c$ than the one reported in the literature (2.5 K)[11] is consistent with the high quality of the as-grown $CsV_3Sb_5$ sample, which provides the opportunity for deeper understanding of the nature of superconductivity and coexisting electronic order through atomically-resolved STM/S. In the STM measurements at 4.2 K, we observe both types of the surfaces (Figs. 1**c-d**). Based on the atomically-resolved STM image combined with crystal structure, we identify the two cleaved terminations as $\sqrt{3}\times\sqrt{3}R30°$ reconstructed Cs surface and 1×1 Sb surface (Methods and Extended Data Fig. 1). In this work, the STM/S imaging of the density waves is conducted on large and clean areas of the Sb surfaces (Fig. 1**e**).

## Unconventional strong-coupling superconductor

We first study the SC ground state using the STS described in Methods at an electron temperature 300 mK (see calibration of electron temperatures in Fig. S3). We observe particle-hole symmetric differential conductance ($dI/dV$) near the Fermi level ($E_F$). The spatially-averaged $dI/dV$ spectra (Fig. 2**a**) show the V-shaped gap on both Cs and Sb surfaces with two gap-edge peaks at energies symmetric with respect to $E_F$. The V-shaped gap is consistent with the gap nodes seen in the thermal transport[19], but the non-zero local density of states (LDOS) at zero-bias, lower on the Sb than the Cs surface (Fig. S4), indicates additional in-gap quasiparticle states possibly due to line nodes or ungapped Fermi surface sections. From the $dI/dV$ spectra collected over nearly fifty 30 nm×30 nm regions, we obtain the average SC gap size ($\Delta$) ~ 0.52±0.1 meV (Fig. S5). The temperature evolution of the $dI/dV$ spectra on the Cs surface (Fig. 2**b**) shows that the V-shaped gap reduces with increasing electron temperature and vanishes around ~ 2.3 K.

Since the suppression of the LDOS near $E_F$ can arise from physics other than superconductivity[20,21], it is crucial to directly probe the superfluid for superconducting phase coherence. We thus construct a Josephson STM (Fig. 2**c**) by fabricating a SC Nb STM tip (see Methods and Fig. S6). A sharp zero-bias peak with two negative differential conductance dips are observed due to Josephson tunneling of Cooper pairs (Fig. 2**d**), which provides strong evidence that the $CsV_3Sb_5$ surface is in the SC phase (Fig. S7). The temperature dependent d$I$/d$V$ spectra are then obtained (Figs. 2**e-f**). When the electron temperature increases to about 900 mK, two sets of peaks can be clearly resolved with particle-hole symmetry. The outer peaks correspond to the sum $\Delta_{tip}+\Delta_{sample}$ of the paring gaps in the tip and the sample, while the inner peaks relate to the difference $\Delta_{tip}-\Delta_{sample}$. Upon further increasing the temperature, the inner peaks disappear at a transition temperature of ~2.3 K, leaving the two remaining peaks from $\Delta_{tip}$. Since the $T_c$ of the Nb tip is higher than 4.2 K (Fig. S6), the transition indicates that superconductivity in the sample is completely suppressed at this temperature. The SC gap of the sample is in excellent agreement with the value measured by the normal W tip. We thus conclude that the observed V-shaped gap (Fig. 2**a**) is the SC gap that opens at a $T_c$ ~ 2.3 K on the surface of the $CsV_3Sb_5$. The V-shaped SC gap with non-zero LDOS at $E_F$ is strongly indicative of unconventional superconductivity[22,23]. Moreover, the measured gap–to–$T_c$ ratio $2\Delta/k_BT_c \sim 5.2$ puts the unconventional superconductor in the strong-coupling regime.

**Coexisting charge density waves**

We next probe the spatial distribution of the off-diagonal long-range ordered quantum states using the high-resolution STM/STS at an electron temperature 300 mK well below $T_c$. Since the Cs atoms are unstable on the Cs terminated surface and strongly affect the tip states, we perform the measurements on the Sb surface directly above the V kagome plane (Fig. 1**e**). The STM topography (Fig. 3**a**) over a large 70 nm×70 nm area (Methods) shows periodic modulations indicating the presence of CDWs. The corresponding Fourier transform after the Lawler-Fujita drift-correction[24,25] (Methods) reveals in Fig. 3**b** that, in addition to the atomic Bragg peaks $\mathbf{Q}^{(a,b)}_{\mathbf{Bragg}}$ of the pristine Sb lattice, two sets of new peaks. One set comprises six hexagonal wave vectors at $\mathbf{Q}_{\mathbf{3q\text{-}2}a} = \frac{1}{2}\mathbf{Q}^{(a,b)}_{\mathbf{Bragg}}$, corresponding to a $2a_0\times2a_0$ superstructure on the kagome lattice. The other set is at uniaxial wave vectors marked by $\mathbf{Q}_{\mathbf{1q\text{-}4}a} = \frac{1}{4}\mathbf{Q}^{a}_{\mathbf{Bragg}}$, corresponding to unidirectional $4a_0$ modulations. These robust modulations are clearly visible in the topography (Fig. 3**a**, inset) and suggest the coexistence of superconductivity with $2a_0$ bidirectional and $4a_0$ unidirectional positional and rotational order, i.e. a smectic state in the language of liquid crystals[26,27]. The d$I$/d$V$($\mathbf{r}$,$V$) conductance map at V=-5 mV over the same field of view (Fig. 3**c**) and the corresponding Lawler-Fujita drift corrected Fourier transform (Fig. 3**d**) reveals the outstanding $\mathbf{Q}_{3q\text{-}2a}$ and $\mathbf{Q}_{1q\text{-}4a}$ peaks and the quasiparticle interference (QPI) patterns. The Bogoliubov QPI patterns in the SC state are notably different, but with certain unidirectional features similar to the normal state QPI[14]. Acquiring additional d$I$/d$V$ maps and Fourier transforms at different bias energies and over different regions, two examples are shown in Figs. 3**e**-**f** around the SC gap energy, we find that the two sets of peaks at $\mathbf{Q}_{3q\text{-}2a}$ and $\mathbf{Q}_{1q\text{-}4a}$ in d$I$/d$V$ maps are nondispersive in energy (Fig. 3**g**, Extended Data Fig. 2 and Fig. S8). We thus conclude that $2a_0 \times 2a_0$ and $4a_0$ unidirectional CDWs[14] coexists with superconductivity, giving rise to the smectic superconductor.

## 3Q PDW with $4a_0/3$ period

A striking feature of the d$I$/d$V$ maps, *absence* in the topography (Fig. 3**b**), is the new set of prominent peaks at hexagonal wave vectors $\mathbf{Q}_{3q\text{-}4a/3} = \frac{3}{4}\mathbf{Q}_{\mathbf{Bragg}}^{(a,b)}$ in the Fourier transform (Figs. 3**d**-**f**). Note that the 3Q peaks at $\mathbf{Q}_{3q\text{-}4a/3}$ are distinct from the uniaxial $\mathbf{Q}_{1q\text{-}4a}$, despite the overlap with a higher harmonic of $\mathbf{Q}_{1q\text{-}4a}$ along the $q_a$-direction. These peaks do not exist in the d$I$/d$V$ maps at energies much higher than the SC gap energy (Extended Data Fig. 3). They suggest an emergent $4a_0/3$ bidirectional 3Q electronic modulations without long-range charge order. The $\mathbf{Q}_{3q\text{-}4a/3}$ peaks are nondispersive at low energies around the SC gap (Fig. 3**g**), indicating the possible formation of a primary 3Q PDW. This is in contrast to the long-range CDWs at $\mathbf{Q}_{3q\text{-}2a}$ and $\mathbf{Q}_{1q\text{-}4a}$ that induce subsidiary PDWs at the same wave vectors in the SC condensate[28-34]. The d$I$/d$V$ ($\mathbf{r}$,0.4 mV) map after Fourier filtering of atomic Bragg peaks at $\mathbf{Q}_{\mathbf{Bragg}}$ and noise from the small-$q$ quasiparticle scattering (Fig. S9), reveals the spatial pattern of the PDW (Fig. 3**h**). A hexagonal pattern of the bidirectional PDW with period $4a_0/3$ (Fig. S9) is clearly observed in the background of the $4a_0$ charge stripes.

## Spatial modulations of SC gap, coherence peak height, and gap depth at PDW wave vector

To investigate the properties of the PDW and its effects on superconductivity, we measure the d$I$/d$V$($\mathbf{r}$,$V$) spectra along a line cut parallel to the $4a_0$ stripes indicated in the topography (Fig. 3**i**). This direction corresponds to the $\mathbf{q}_b$-direction in reciprocal space (Fig. 3**b**), thus avoids the $4a_0$ modulations due to the unidirectional charge order. The spatial evolution of the differential conductance (Fig. 3**j**) displays intricate modulations in the SC gap $\Delta(\mathbf{r})$, the coherence peak height at the gap edge, and the zero-bias conductance. To extract quantitative information, we take the second derivative of each conductance curve: $D(\mathbf{r},V)= -\mathrm{d}^3I/\mathrm{d}V^3(\mathbf{r},V)$. The peaks in $D(\mathbf{r},V)$ along the cut (Fig. 3**k**) determines accurately the coherence peak locations (especially at negative bias) and the SC gap $\Delta(\mathbf{r})$, similar to a recent study of the PDW modulation of the SC coherence in underdoped cuprates[29]. The spatial modulation of the local SC gap $\Delta(\mathbf{r})$ (Fig. 3**l**), having an amplitude on the order of 7% of the average gap, is clearly visible with underlying periodicities. Its Fourier spectrum (Fig. 3**l**) displays pronounced peaks at $\frac{3}{4}\mathbf{Q}_{\mathbf{Bragg}}$ associated with the PDW, in addition to the Bragg peak at $\mathbf{Q}_{\mathbf{Bragg}}$ and the $2a_0$ CDW peak at $\frac{1}{2}\mathbf{Q}_{\mathbf{Bragg}}$. The modulation of the SC gap further supports the identification of the $\mathbf{Q}_{\mathbf{3q\text{-}4}a\mathbf{/3}}$ peaks in the d$I$/d$V$ maps (Figs. 3**e**,**f**) with the 3Q PDW coupled to the SC condensate. More data, including the SC gap maps acquired in different regions exhibiting $\mathbf{Q}_{\mathbf{3q\text{-}4}a\mathbf{/3}}$ peaks in the Fourier transforms, are presented in Figs. S10 and S11.

To demonstrate the modulation of the SC coherence by the PDW, we determine from the conductance spectrum (Fig. 3**j**) the coherence peak height at the gap edge: $P(\mathbf{r}) = \mathrm{d}I/\mathrm{d}V(\mathbf{r},+\Delta(\mathbf{r}))$, the zero-bias conductance $G_0(r)=\mathrm{d}I/\mathrm{d}V(\mathbf{r},0)$, and the gap-depth $H(\mathbf{r})=P(\mathbf{r})-G_0(\mathbf{r})$. The raw data for $P(\mathbf{r})$, $H(\mathbf{r})$, and $G_0(\mathbf{r})$ (Fig. 3**m**) display intriguing periodic modulations along the line cut. Remarkably, the modulations of the coherence peak $P(\mathbf{r})$ and zero-bias conductance $G_0(\mathbf{r})$ are out of phase (Figs. 3**m**, S12**a**), leading to in-phase modulations of coherence peak $P(\mathbf{r})$ and gap-depth $H(\mathbf{r})$. Since a higher coherence peak in STS usually reflects a higher superfluid density[29], the correlated modulations among $P(\mathbf{r})$, $H(\mathbf{r})$, and $G_0(\mathbf{r})$ amount to a concomitant deeper SC gap and less normal fluid density, demonstrating an unprecedented electronic density wave modulation of superconductivity. After filtering out atomic Bragg oscillations and small-$q$ noise due to quasiparticle scattering from the raw data, the spatial modulation of the coherence peak height $P(\mathbf{r})$ and the SC gap-depth $H(\mathbf{r})$ exhibit remarkable beating patterns of two primary frequencies corresponding to the leading bidirectional $4a_0/3$ PDW and a weaker $2a_0$ CDW (Fig. 3**n** and Fig. S12**b**). These results demonstrate that the emergent PDW involves both the superfluid and the normal fluid, such that the total electron density in the ground state is only weakly perturbed at the bidirectional $\mathbf{Q}_{\mathbf{3q\text{-}4}a\mathbf{/3}}$.

The observation of the PDW at $\mathbf{Q}_{\mathbf{pdw}} = \mathbf{Q}_{\mathbf{3q\text{-}4}a/3}$ in the superconductor is consistent with the existence of a shallow roton minimum at same wave vector $\mathbf{Q}_{\mathbf{roton}} = \mathbf{Q}_{\mathbf{3q\text{-}4}a/3}$ in the dynamical density-density response function of the superfluid (Extended Data Fig. 4). The roton gap protects the superfluid from crystallization, allowing only short-ranged charge density correlations. The 3Q PDW is described by an inhomogeneous order parameter $\Delta_{\text{pdw}}(\boldsymbol{r}) = \sum_\alpha \Delta_{\boldsymbol{Q}_\text{p}^\alpha}(\boldsymbol{r})$, where $\Delta_{\boldsymbol{Q}_\text{p}^\alpha}(\boldsymbol{r}) = \Delta_\alpha \cos(\boldsymbol{Q}_\text{p}^\alpha \cdot \boldsymbol{r} - \phi_\alpha)$ and $\boldsymbol{Q}_\text{p}^\alpha$, $\alpha = 1, 2, 3$, are the momenta corresponding to $\mathbf{Q}_{\mathbf{3q\text{-}4}a/3}$ in the 3Q directions and $\phi_\alpha$ is a relative phase. Below $T_c$, it couples to the uniform SC condensate $\Delta_{sc}$. Thus, the intertwined density wave order has the character of delocalized Cooper pair excitations and localized charge excitations. To stress this distinction, we refer to this novel quantum state as a roton-PDW, without implying direct observation of roton excitations, which may be visible in the conductance spectrum at higher energies through mode coupling. In this scenario, the low energy excitations involve both gapless quasiparticles and roton-PDW excitations. Since a roton is a bound vortex-antivortex pair[35-38], the roton-PDW can be viewed as a commensurate hexagonal vortex-antivortex lattice (Extended Data Fig. 4) from the zeroes in the complex PDW order parameter $\Delta_{\text{pdw}}(\mathbf{r})$ that coexists with the uniform component of the SC order parameter $\Delta_{sc}$. Such an unconventional SC state necessarily breaks the time-reversal symmetry, exhibiting spontaneous phase windings associated with $\Delta_{\text{pdw}}(\mathbf{r})$ and unconventional SC vortices[30,33]. The roton-PDW provides a qualitative explanation for our observations. The low energy states inside the V-shaped SC gap (Fig. 2**a**) are composed of both localized vortex-antivortex core states and itinerant nodal quasiparticles contributing to the observed thermal transport[19]. The spatial modulation of the coherence peak height and zero-bias conductance can be accounted for by those in the local superfluid density and the anti-correlated normal fluid density on the vortex-antivortex lattice (Extended Data Fig. 4) under weak modulations of the SC gap.

### PDW as a "mother state" and emergent pseudogap

We further investigate the nature of the PDW by applying a magnetic field along the c-axis. The magnetic field dependent d*I*/d*V* spectra on the Sb surface away from field-induced vortices show that the SC gap is gradually reduced with increasing field and vanishes at about 2 T (Fig. S13). At 0.04 T, we observed SC vortices (Fig. S14). Fig. 4**a** centers on a vortex in the hexagonal vortex lattice. In the vortex halo marked in Fig. 4**a,** we obtain the d*I*/d*V*(**r**,-5 mV) map (Fig. 4**b**). The Fourier transform shows that the 3Q PDW at $\mathbf{Q}_{3q\text{-}4a/3}$ survives in the vortex halo, with somewhat split peaks (Fig. 4**c**). In contrast to the cuprates, where the $8a_0$ PDW appears only in the vortex halo with suppressed but nonzero superconductivity[33,39], the $4a_0/3$ roton-PDW is strong enough to emerge both in the vortex halo and in the full-fledged superconductor in zero-field. When the magnetic field is raised to 2 T, the SC gap disappears and superconductivity is suppressed at 300 mK (Fig. S13). The d$I$/d$V$(**r**,-5 mV) map and the Fourier transform (Fig. 4**d**-**e**) show that, while the QPI pattern changes substantially, all the density wave peaks remain including the $\mathbf{Q}_{3q\text{-}4a/3}$ associated with the $4a_0/3$ PDW. With superconductivity removed by the magnetic-field, these coexisting density waves define a "zero-temperature" pseudogap phase with a suppression of the LDOS over ±5 meV in the spatially-averaged d$I$/d$V$ spectrum shown in Fig. 4**h**. In contrast to the 1×4 and 2×2 CDW peaks that exist at all energies, the nondispersive $\mathbf{Q}_{3q\text{-}4a/3}$ peaks are visible only in the energy range of the pseudogap (~ ±5 meV) and are absent at higher energies (Extended Data Fig. 3). This suggests an intriguing possibility that the observed $4a_0/3$ 3Q PDW is a "mother state" responsible for the pseudogap, as proposed in theories of the cuprates[32,40]. A Gizsburg-Landau analysis (Methods 'Discussions of PDW as a 'mother state'') shows that due to the hexagonal symmetry, the 3Q PDW induces a secondary 3Q CDW of identical period, giving rise to an intertwined electronic order at $\mathbf{Q}_{3q\text{-}4a/3}$ as the "mother state" responsible for the pseudogap. We have identified the PDW pseudogap with the energy of the peak in the LDOS near 5 mV in the SC state at 300 mK (Methods and Extended Data Fig. 5) and acquired the pseudogap map, which indeed exhibits spatial gap modulations with Fourier peaks at the PDW vector $\mathbf{Q}_{3q\text{-}4a/3}$.

Finally, we warm up the sample to the normal state above $T_c$ and acquired d$I$/d$V$ maps (Fig. 4**f**) in the same region at 4.2 K. The corresponding Fourier transform (Fig. 4**g**) shows modified QPI patterns from the SC state at 300 mK (Fig. 3**d**) due to the closing of the SC gap. The PDW peaks at $\mathbf{Q}_{3q\text{-}4a/3}$ are somewhat more diffused, but clearly present, demonstrating that the PDW and the intertwined electronic order persist to the normal state. The spatially-averaged d$I$/d$V$ spectrum at 4.2 K (Fig. 4**h**) indeed exhibits a broad incoherent normal state pseudogap correlated with the primary PDW over the energy range ~±5 meV. The comparison of the averaged d$I$/d$V$ spectra (Fig. 4**h**) obtained in different states including the SC state at 0 T, the vortex halo at 0.04 T, the "zero-temperature" pseudogap state at 2 T and 300 mK, and the normal state pseudogap phase at 4.2 K, reveals a rather consistent and enlightening picture of the interplay between superconductivity, the primary PDW, the intertwined density waves and the pseudogap with striking analogy and distinction to the physics of the high-$T_c$ cuprates.

The d$I$/d$V$ map and Fourier transform (Fig. 4**f**-**g**) indicate $2a_0\times2a_0$ CDW and $4a_0$ unidirectional charge order persist above the SC transition. The properties of the CDWs in the normal state, discussed in more detail in Methods together with our DFT calculations (Extended Data Fig. 6 and Figs. S15, S16), are in good agreement with the recent STM work[14].The angular-dependent magnetoresistance measurements reveal a twofold resistivity anisotropy (Extended Data Fig. 7) with a sharp onset below ~50 K, which matches well with the onset temperature $T_{\mathrm{stripe}}$~50 K of the $4a_0$ stripes detected by STM[14]. This suggests an incipient rotational symmetry breaking bulk electronic state below $T_{\mathrm{stripe}}$, which can be either a quasi-3D $4a_0$ stripe phase with interlayer coupling, or a different state that manifests as the $4a_0$ unidirectional charge order on the Sb terminated surface of $CsV_3Sb_5$.

Unconventional superconductivity can arise in model calculations from local and extended electron correlations on the kagome lattice[41-43]. We stress, however, that the physics discovered here goes well beyond $CsV_3Sb_5$ being another candidate unconventional superconductor. It embodies a set of highly provoking quantum electronic states and excitations that show striking analogies and distinctions and may hold the common set of keys to resolve some of the outstanding issues in the cuprate high-$T_c$ superconductors, including smectic electronic liquid crystal states, the interplay among PDW, CDW and intertwined electronic order, as well as their impact on the pseudogap phenomenon and unconventional superconductivity. Our findings provide groundwork and insights for future studies on how the unconventional SC state, the roton-PDW, and the coexisting charge order originate microscopically from the correlated $Z_2$ topological kagome bands, and on the prospects of emergent topological superconductivity in $AV_3Sb_5$.

## Main Figures

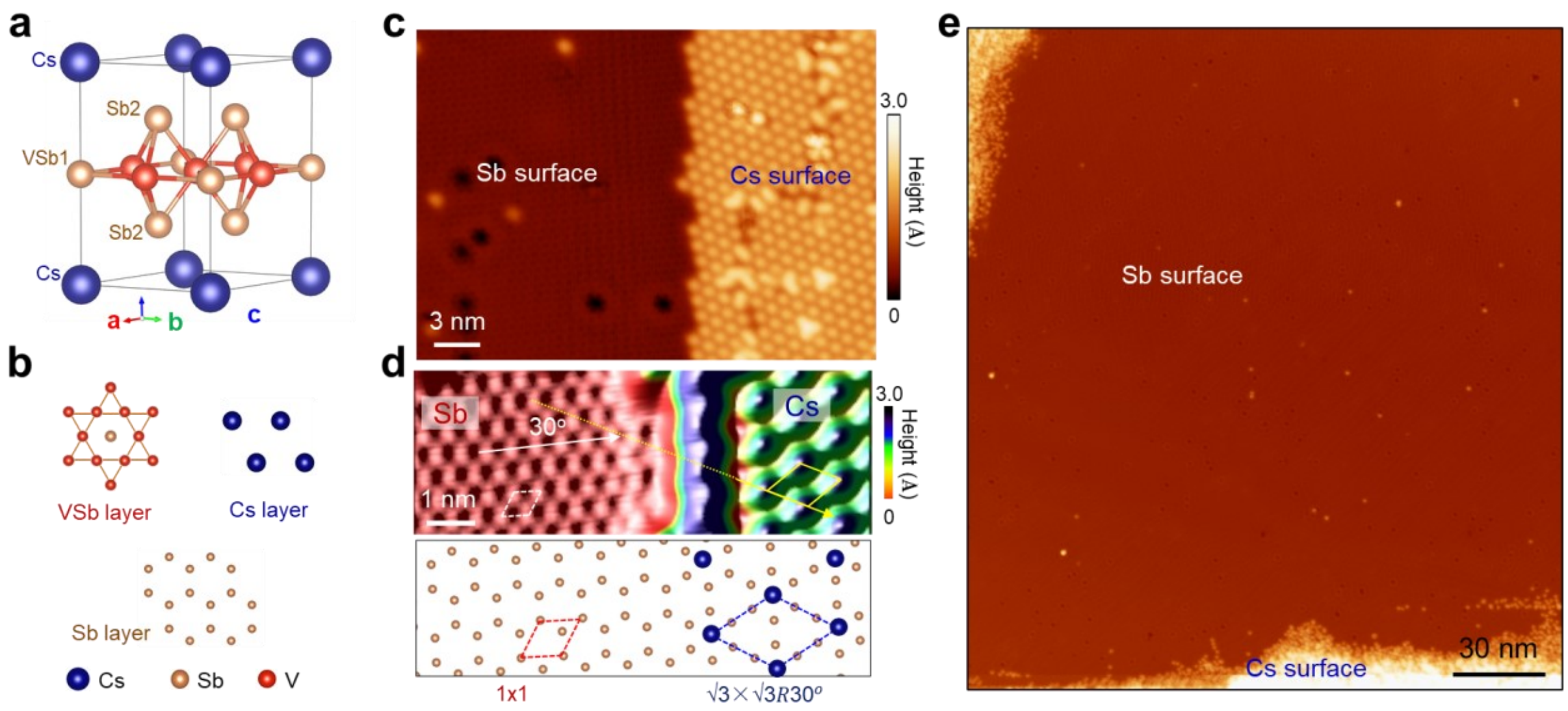

**Fig. 1. Atomic structures and the surface identification of the $CsV_3Sb_5$. a,** 3D view of crystal structure, showing stacking of Cs-Sb2-VSb1-Sb2-Cs layers with hexagonal symmetry. **b**, Atomic structures of VSb, Cs and Sb layers. **c**, Typical large-scale STM image, showing the Cs surface and Sb surface, respectively (scanning setting: bias: $V_s$=-1.0 V, setpoint $I_t$=100 pA). **c,** Atomically-resolved STM image, showing the atomic structures of the Cs surface and the Sb surface, respectively ($V_s$=-1.0 V, $I_t$=100 pA). **d.** Zoom-in of (d) showing the lattice orientation between the Cs surface and Sb surface ($V_s$=-0.5 V, $I_t$=500 pA). Bottom panel: schematic atomic structures of the Sb and Cs surfaces, showing 1×1 and $\sqrt{3}\times\sqrt{3}R30^o$ reconstructed structures, respectively. **e,** STM image of a large and clean Sb surface ($V_s$=-2.0 V, $I_t$=100 pA).

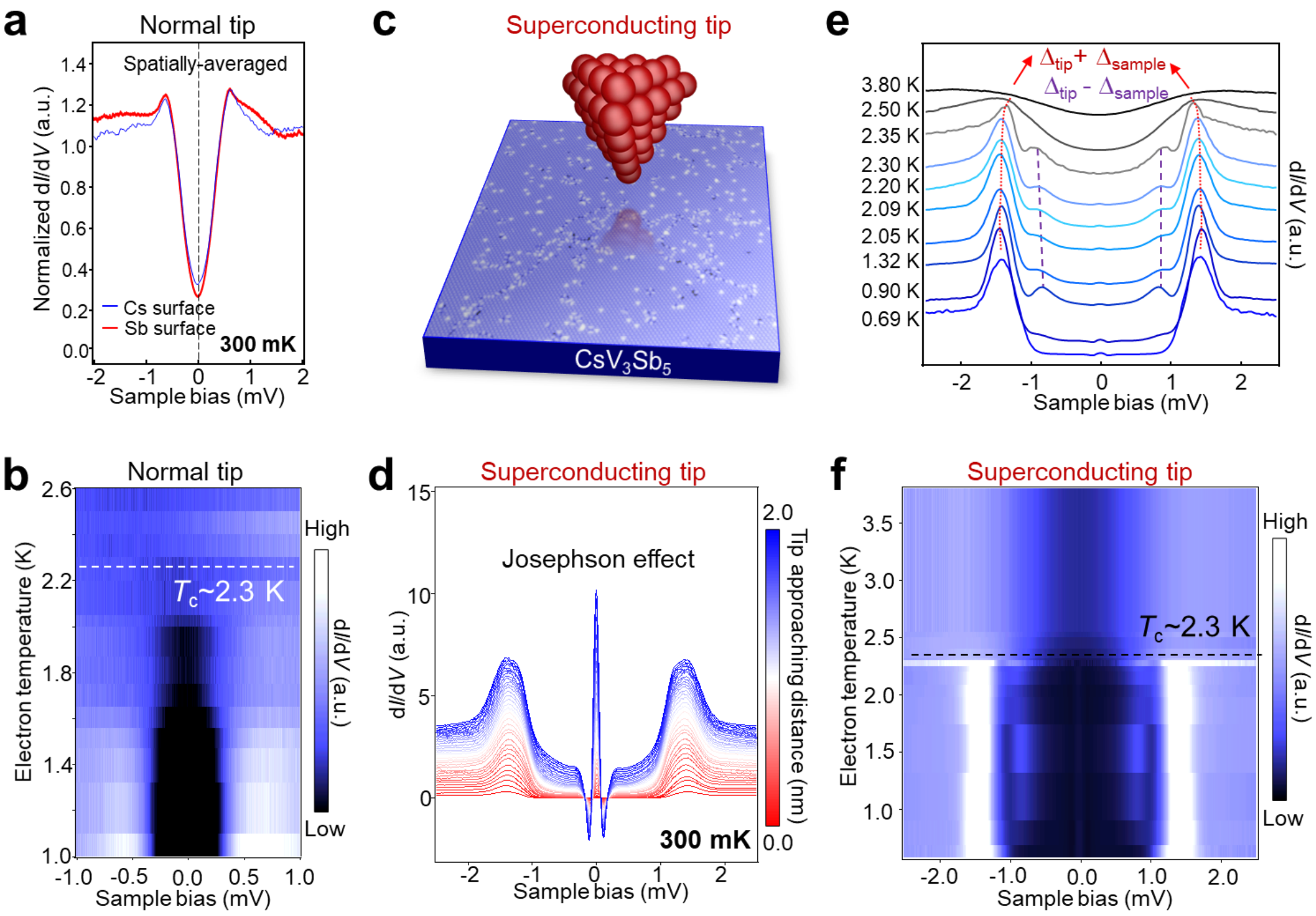

**Fig. 2. V-shaped pairing gap and the Josephson effect observed using a superconducting STM tip on the Cs and Sb surfaces. a,** Spatially-averaged d$I$/d$V$ spectra obtained on the Cs and Sb surfaces over a 30 nm×30 nm region with small defect density at 300 mK, showing a particle-hole symmetric V-shaped gap near $E_F$ and non-zero LDOS at zero-bias ($V_s$=-2 mV, $I_t$=1 nA, $V_{mod}$=50 μV). **b**, Color map of temperature dependent d$I$/d$V$ spectra obtained on the Cs surface, showing that the V-shaped gap reduces with increasing electron temperature and vanishes around ~ 2.3 K (dotted white line) ($V_s$=-2 mV, $I_t$=1 nA, $V_{mod}$=50 μV). **c,** Schematics showing the Josephson STM on $CsV_3Sb_5$ surface using a superconducting (Nb) tip. **d**, A series of d$I$/d$V$ spectra with approaching tip toward sample surface. A sharp zero-bias peak with two negative differential conductance dips are observed, providing strong evidence that the $CsV_3Sb_5$ is in the SC phase ($V_s$=-2.5 mV, $V_{mod}$=50 μV). **e,** Temperature dependent d$I$/d$V$ spectra on the Cs surface using a Nb tip at a relatively small tip-sample distance where the zero-bias peak is still present, showing two energy gaps at $\Delta_{tip}+\Delta_{sample}$ and $\Delta_{tip}-\Delta_{sample}$, respectively ($V_s$=-2.5 mV, $I_t$=20 nA, $V_{mod}$=50 μV ). **f**, Color map of the d$I$/d$V$ spectra in (e), showing the transition temperature of the $CsV_3Sb_5$ sample around ~2.3 K ($V_s$=-2.5 mV, $I_t$=20 nA, $V_{mod}$=50 μV).

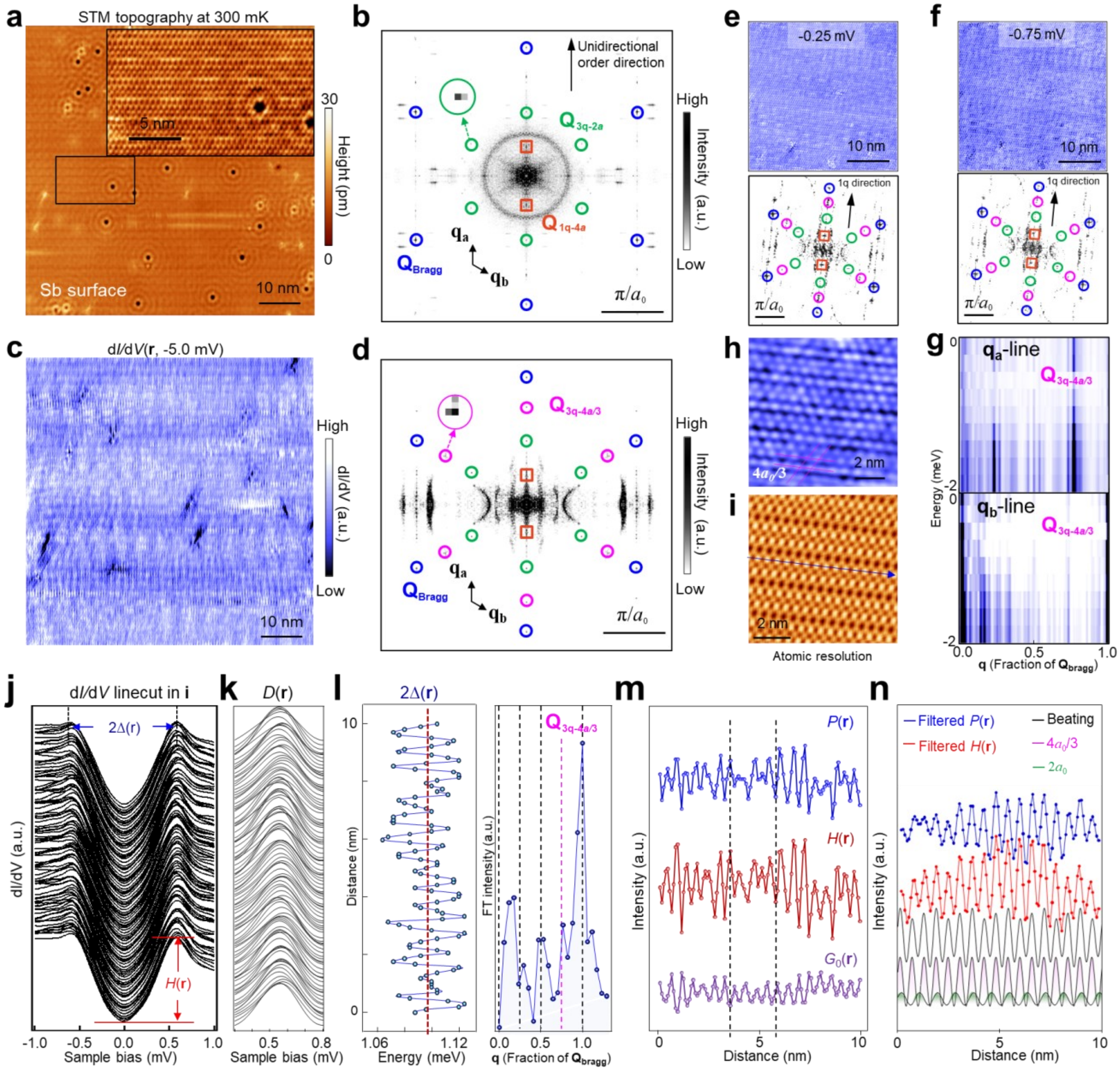

**Fig. 3. STM topography, d*I*/d*V* map and line cut at 300 mK revealing CDW, PDW, and spatial modulations of superconductivity on Sb surfaces. a,b** The large-scale STM topography of Sb surface and the magnitude of drift-corrected, two-fold symmetrized Fourier transform (FT), showing wave vectors $\mathbf{Q}_{3q\text{-}2a}$ for 2×2 CDW and $\mathbf{Q}_{1q\text{-}4a}$ for $4a_0$ unidirectional charge order ($V_s$=-10 mV, $I_t$=500 pA ). Inset: Zoom-in of (a), exhibiting atomically-resolved STM image ($V_s$=-5 mV, $I_t$=1 nA). **c,** d*I*/d*V*(**r**,-5 mV) map and the magnitude of drift-corrected, two-fold symmetrized FT, revealing new 3Q PDW modulations at $\mathbf{Q}_{3q\text{-}4a/3}$ ($V_s$=-5 mV, $I_t$=1 nA, $V_{mod}$=0.2 mV). **e, f,** d*I*/d*V*(**r**,-0.25 mV), d*I*/d*V*(**r**,-0.75 mV) maps and the magnitude of drift-corrected FTs over a different region from (a). Besides the $\mathbf{Q}_{3q\text{-}2a}$ and $\mathbf{Q}_{1q\text{-}4a}$ charge ordered states present in the topography, the new 3Q PDW modulations at $\mathbf{Q}_{3q\text{-}4a/3}$ can also be observed ($V_s$=-5 mV, $I_t$=1 nA, $V_{mod}$=0.2 mV). **g,** Energy dependence of the Fourier line cuts along the $\mathbf{q}_a$-direction and $\mathbf{q}_b$-direction as a function of energy, respectively, showing non-dispersive ordering vectors at $\mathbf{Q}_{3q\text{-}4a/3}$. **h**, d*I*/d*V*(*r*,*V*) map near the pairing gap at 0.4 mV, after filtering out atomic Bragg peaks and incoherent background, exhibiting $4a_0/3$ checkerboard modulation associated with the bidirectional PDW. **i**. Atomically-resolved STM image displaying the $\mathbf{Q}_{1q\text{-}4a}$ spatial modulation ($V_s$=-90 mV, $I_t$=2 nA). **j,k** Evolution of the differential conductance d*I*/d*V*(**r**,*V*) and negative of its second derivative $D(\mathbf{r},V)=-d^3I/dV^3(\mathbf{r},V)$ along the linecut in $\mathbf{q}_b$-direction marked by the blue arrow in (i), respectively. ($V_s$=-1 mV, $I_t$=1 nA, $V_{mod}$=50 μV). **l,** Plot of 2Δ(**r**) and the corresponding Fourier spectrum, showing the $4a_0/3$ spatial modulation highlighted by the pink dashed line marked by $Q_{3q\text{-}4a/3}$. **m**, Spatial modulations of coherence peak height *P*(**r**), SC gap depth *H*(**r**), and zero-bias conductance $G_0$(**r**). The spectra are offset for clarity. **n**, Beating pattern simulated using $\mathbf{Q}_{3q\text{-}2a}$ (green curve) and $\mathbf{Q}_{3q\text{-}4a/3}$ (pink curve), showing consistency with Bragg-filtered modulations of *P*(**r**) and *H*(**r**).

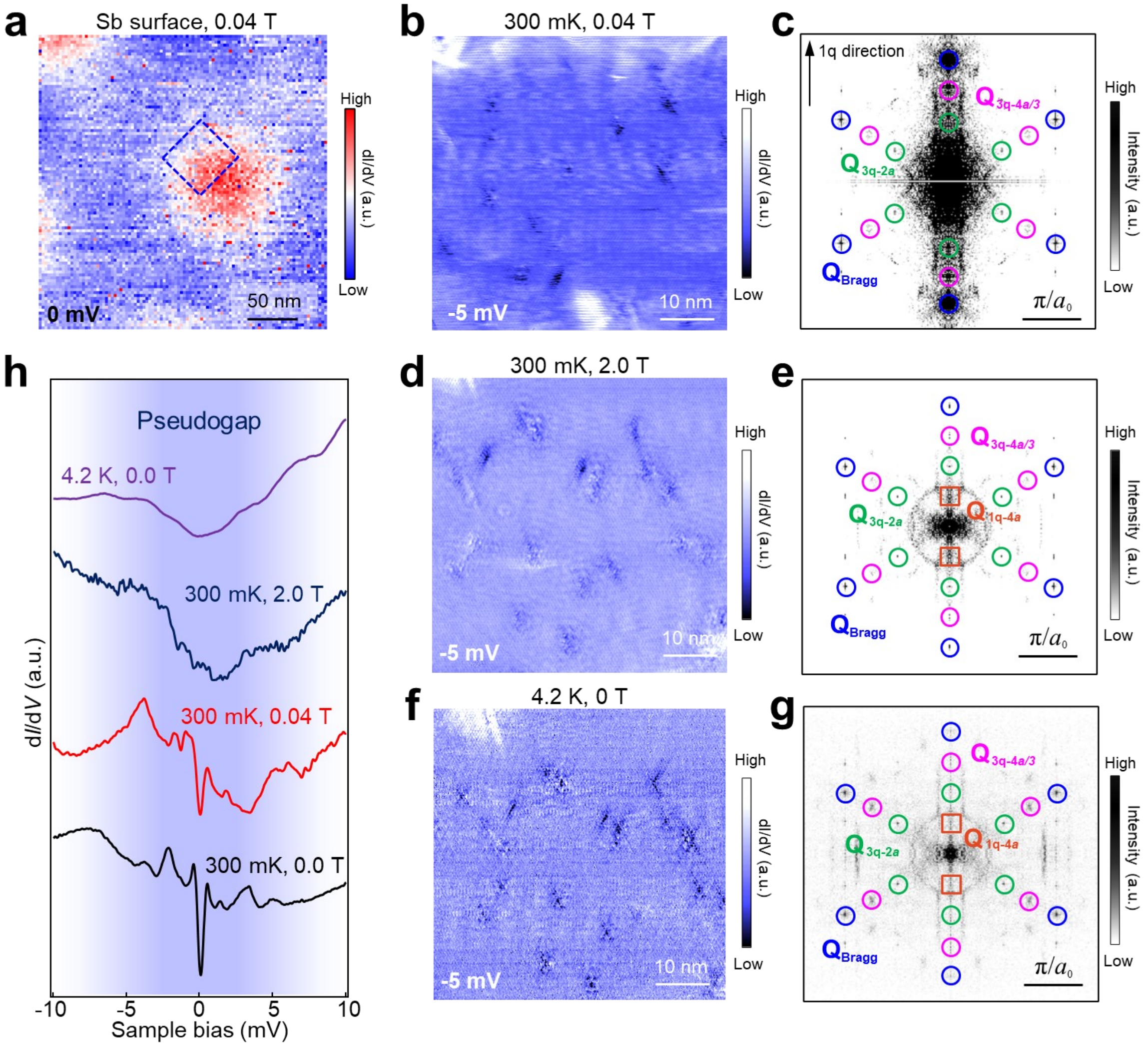

**Fig. 4. PDW and pseudogap in CsV3Sb5 in magnetic fields at 300 mK and in zero-filed at 4.2 K. a**, d*I*/d*V(r,*0*)* map of Sb surface in a 0.04 T magnetic field at 300 mK, showing the vortex lattice ($V_s$=-1 mV, $I_t$=1 nA, $V_{mod}$=50 µV). **b,** d*I*/d*V*(**r***,* -5 mV) map in vortex halo marked by the blue square in (a) ($V_s$=-5 mV, $I_t$=1 nA, $V_{mod}$=0.2 mV). **c**, The magnitude of drift-corrected, two-fold symmetrized Fourier transform (FT) of (b). **d,e** d*I*/d*V*(**r***,* -5 mV) map and the magnitude of drift-corrected, two-fold symmetrized FT in a 2.0 T at 300 mK ($V_s$=-5 mV, $I_t$=1 nA, $V_{mod}$=0.2 mV). **f,g** d*I*/d*V*(**r***,* -5 mV) map under zero-field at 4.2 K and the magnitude of drift-corrected, two-fold symmetrized FT ($V_s$=-5 mV, $I_t$=1 nA, $V_{mod}$=0.2 mV). **h**, Spatially-averaged d*I*/d*V* spectra over the region marked in **a** obtained under different physical conditions of magnetic field and temperature, showing the spectral changes and the presence of the low energy pseudogap ($V_s$=-10 mV, $I_t$=1 nA, $V_{mod}$=50 µV). The spectra are offset for clarity.

## Methods

**Single crystal growth of $CsV_3Sb_5$ sample**. Single crystals of $CsV_3Sb_5$ were grown from Cs liquid (purity 99.98%), V powder (purity 99.9%) and Sb shot (purity 99.999%) via a modified self-flux method[10]. The mixture was put into an alumina crucible and sealed in a quartz ampoule under Argon atmosphere. The mixture was heated to 1000 ºC and soaked for 24 h, and subsequently cooled at 2 ºC/h. Finally, the single crystal was separated from the flux and the residual flux on the surface was carefully removed by a Scotch tape. Except sealing and heat treatment procedures, all of other preparation procedures were carried out in an argon-filled glove box in order to avoid the introduction of any air and water. The obtained crystals have a typical hexagonal morphology with a size of over 2×2×0.3 $cm^3$ (Fig. S1) and are stable in the air.

**Sample characterization**. XRD pattern was collected using a Rigaku SmartLab SE X-ray diffractometer with Cu Kα radiation (λ = 0.15418 nm) at room temperature. Scanning electron microscopy (SEM) and energy-dispersive X-ray spectroscopy (EDX) were performed using a HITACHI S5000 with an energy dispersive analysis system Bruker XFlash 6|60. Magnetic susceptibility were determined by a SQUID magnetometer (Quantum Design MPMS XL-1). The SC transition of each sample was monitored down to 2 K under an external magnetic field of 1 Oe. Both in-plane electrical resistivity and Hall resistivity data were collected on a Quantum Design Physical Properties Measurement System (PPMS).

**Surface determination**. The weak bonds between Cs and Sb2 layers offer a cleave plane and make it possible to have both Cs-terminated and Sb-terminated surfaces. At the interface of the two surfaces, we can clearly identify the atomic structures on both the top and bottom surfaces using high-resolution STM. We find that the lattice in the bottom layer shows a honeycomb configuration, which matches that of Sb2 layer, while the top surface shows a hexagonal lattice with a spacing of 1 nm, which is about √3 larger than the lattice constant of the pristine Cs surface (Extended Data Fig. 1).

**Scanning tunneling microscopy/spectroscopy.** The samples used in the experiments are cleaved at room temperature (300 K) or low temperature (78 K) and immediately transferred to an STM chamber. Experiments were performed in an ultrahigh vacuum ($1\times10^{-10}$ mbar) ultra-low temperature STM system equipped with 9-2-2 T magnetic field. The electronic temperature in the low-temperature STS is calibrated (Fig. S3) using a standard superconductor, Nb crystal. All the scanning parameters (setpoint voltage and current) of the STM topographic images are listed in the figure captions. Unless otherwise noted, the d$I$/d$V$ spectra were acquired by a standard lock-in amplifier at a modulation frequency of 973.1 Hz. Non-superconducting tungsten tips were fabricated via electrochemical etching and calibrated on a clean Au(111) surface prepared by repeated cycles of sputtering with argon ions and annealing at 500 °C.

**Josephson Scanning spectroscopy.** If the sample surface is SC, a superconductor-insulator-superconductor (SIS) junction naturally forms under a SC STM tip for SIS tunneling. This is known as a Josephson STM. In this case, the SC tip and sample are coupled by the Josephson coupling $E_J$ as the tip approaches sufficiently close to the sample at temperatures below the SC transition temperatures of both the tip and sample. At temperatures $k_B T > E_J$, thermally excited Cooper pairs tunnel and give rise to a tunneling current directly proportional to the phase difference between the SC order parameters in the tip and the sample. This Josephson effect – a sharp zero-bias peak accompanied by negative differential conductance dips on both sides, is a phase-sensitive probe for the existence of superconductivity in the sample[44-46]. We use a SC Nb tip in the Josephson STM measurement (Fig. S6). The Nb tip was fabricated via mechanical cut of a Nb rod and calibrated on a clean Nb(110) surface prepared by repeated cycles of sputtering with argon ions and annealing at 1200 °C. Based on the observed $\mathrm{d}I/\mathrm{d}V$ spectra associated with the Josephson effect in Fig. 2**d**, we estimate the Josephson energy $E_J = 1.61$ μeV and the charging energy $E_C = 160$ μeV, which allow the thermal energy to satisfy $E_J < E_T < E_C$, where $E_T = k_B T = 25.8$ μeV is the thermal energy at electron temperature 300 mK. Thus, the zero-bias peak in Fig. 2**d** can be well explained by Cooper pair tunneling in the fluctuation dominated Josephson regime[44-50].

**Procedure to obtain large areas of exposed Sb surface.** In our STM measurements, almost all the surface regions in more than ten as-cleaved $CsV_3Sb_5$ samples show large-sized Cs surface topography; the large-sized Sb surface topography is rarely observed. However, we find that the Cs atoms on the Cs surface is weakly coupled to the bottom Sb surface at low temperatures. Using STM tip manipulation, we can successfully 'push' the Cs atoms on the Cs surface away and expose the bottom Sb surface (Extended Data Fig. 1**f-g**), independent of the cleaving temperature. Repeating the procedure hundreds of times, the exposed Sb surface can become as large as 150 nm × 80 nm (Fig. S15).

**Drift correction and two-fold symmetrized Fourier transform.** To eliminate the STM tip-drift effect from the topography and $\mathrm{d}I/\mathrm{d}V$ map, we apply the well-known Lawler-Fujita algorithm[24,25], after subtracting a 2nd or 3rd degree polynomial background, and obtain a set of displacement fields and drift-corrected topography. The so obtained displacement fields are then applied to the simultaneously measured $\mathrm{d}I/\mathrm{d}V$ maps in the same field of view as the topography. To reduce background noise, we also apply the two-fold symmetrization along the axis perpendicular to the $\mathbf{q}_\mathbf{a}$ direction in the Fourier transform in Figs. 3**b**, 3**d** and Figs. 4**c**, 4**e**, 4**g**.

**DFT calculations.** All calculations were performed within the Density Functional Theory as implemented in the Vienna Ab-initio Simulation Package (VASP)[51,52]. The exchange-correlation functional were treated with in the generalized gradient approximation as parametrized by Perdew-Burke-Ernzerhof [53]. The cutoff energy for the plane-wave basis set is 300 eV. The zero damping DFT-D3 van der Waals correction[54] is employed in all the calculations. Spin-orbital coupling (SOC) is considered in the band structure calculations. *k*-meshes of 9×9×6 and 6×6×6 are used for calculating electronic structures of the pristine phase and 2×2 CDW phase, respectively. The phonon dispersions are calculated (without SOC) by using the phonopy code[55] within a 3×3×2 supercell for the pristine structure and 2×2×2 supercells for 2×2 CDW phases.

**Discussions of PDW as a "mother state".** The magnetic field experiments in Fig. 4 raise an intriguing question whether the observed $4a_0/3$ bidirectional PDW is a "mother state" responsible for the pseudogap. The concept of a "mother state" PDW was proposed in theories of the cuprates[32,40], where the $8a_0$ PDW is strong enough to produce a half-period $4a_0$ modulation in the vortex halo, and while fluctuating above $T_c$ induces a secondary $4a_0$ CDW responsible for the pseudogap behavior. We argue that the $4a_0/3$ bidirectional PDW may indeed be a "mother state", but under a different incarnation. In the Ginzburg-Landau theory, in addition to coupling to a SC condensate $\Delta_{sc}$ if available, the products of a robust PDW order parameter with itself can generate an intriguing set of electronic order[30,33,56]. Among them, the induced CDWs are given by $\psi_{2\boldsymbol{Q}_{\mathrm{p}}^{\alpha}}(\mathbf{r}) \propto (\Delta_{\boldsymbol{Q}_{\mathrm{p}}^{\alpha}}(\mathbf{r})\Delta_{-\boldsymbol{Q}_{\mathrm{p}}^{\alpha}}^{*}(\mathbf{r}) + h.c)$ and $\psi_{\boldsymbol{Q}_{\mathrm{p}}^{\alpha}-\boldsymbol{Q}_{\mathrm{p}}^{\beta}}(\mathbf{r}) \propto (\Delta_{\boldsymbol{Q}_{\mathrm{p}}^{\alpha}}(\mathbf{r})\Delta_{\boldsymbol{Q}_{\mathrm{p}}^{\beta}}^{*}(\mathbf{r}) + \Delta_{-\boldsymbol{Q}_{\mathrm{p}}^{\beta}}(\mathbf{r})\Delta_{-\boldsymbol{Q}_{\mathrm{p}}^{\alpha}}^{*}(\mathbf{r}))$, in our notations defined in the main text. In the hexagonal zone, the half-period CDW $\psi_{2\boldsymbol{Q}_{\mathrm{p}}^{\alpha}}$ has wave vectors $2\boldsymbol{Q}_{\mathbf{p}}^{\boldsymbol{\alpha}} = \frac{3}{2}\boldsymbol{Q}_{\mathbf{Bragg}}^{\boldsymbol{\alpha}}$, which are not new but coincide with the existing peaks of the $2a_0\times2a_0$ CDW in the second zone. More surprisingly, because of the hexagonal symmetry, the induced CDW $\psi_{\boldsymbol{Q}_{\mathrm{p}}^{\alpha}-\boldsymbol{Q}_{\mathrm{p}}^{\beta}}$ has a wave vector $\boldsymbol{Q}_{\mathbf{p}}^{\alpha} - \boldsymbol{Q}_{\mathbf{p}}^{\beta} = \boldsymbol{Q}_{\mathbf{p}}^{\gamma}$ , which is identical to one of the three PDW wave vectors $\mathbf{Q}_{3q\text{-}4a/3}$. This remarkable result suggests that the 3Q PDW with period $4a_0/3$ induces a secondary 3Q CDW of identical period, giving rise to an intertwined electronic order at $\mathbf{Q}_{3q\text{-}4a/3}$ as the "mother state" responsible for the pseudogap.

**Spatial modulations of the PDW pseudogap.** We have obtained the Fourier transforms of dI/dV(**r**,-5mV) maps taken at different temperatures and observed that, as the temperature is reduced, the PDW peak intensity at $\mathbf{Q}_{3q\text{-}4a/3}$ normalized by the Bragg peak intensity at $\mathbf{Q}_{\mathbf{Bragg}}$ in the same direction increases substantially below the SC transition temperature $T_c$ ~2.5 K following the onset of the coupling to the SC condensate. This suggests the possibility of identifying the pseudogap energy scale and detecting the pseudogap modulations in the SC state. To this end, we first acquire a spatially-averaged conductance spectrum over a region at 300mK below $T_c$. The spectrum exhibits several peaks in the energy range between 1 mV and 6 mV (Extended data Fig. 5**a**). These peaks are also visible in the corresponding bottom curve in Fig. 4**h**, which are broadened possibly due to averaging over a much larger field of view beyond the coherence length. We find that only the peak located near 5 meV remains prominent above $T_c$ and exhibits periodic spatial modulations as shown in the line cut in Extended data Fig. 5**b**. We thus identify the peak located near 5 meV as the PDW pseudogap peak with its energy position defining the pseudogap size (Δ*) (Extended data Fig. 5**a**). Then, we acquire the pseudogap map Δ*(**r**), which displays spatial modulations of the pseudogap size, having an amplitude on the order of 12% of the averaged pseudogap (Extended data Fig. 5**c**). In the Fourier transform of Δ*(**r**), the peaks at the PDW vector $\mathbf{Q}_{3q\text{-}4a/3}$ can be clearly observed (Extended data Fig. 5**d**). This further demonstrates our observation of the $4a_0/3$ bidirectional PDW as a "mother state" responsible for the pseudogap, which coexists with the SC gap below $T_c$ by coupling to the SC condensed and is detectable by STM.

**Discussions of CDW in normal state**

The d$I$/d$V$ map and Fourier transform (Fig. 4**f**-**g**) indicate long-range $2a_0\times2a_0$ CDW and $4a_0$ unidirectional charge order in the normal state above the SC transition. These can be seen directly from the atomically resolved STM topography over large areas and the d$I$/d$V$ maps measured at different bias-energies with nondispersive $\mathbf{Q}_{3q\text{-}2a}$ and $\mathbf{Q}_{1q\text{-}4a}$ peaks shown in the supplemental (Extended Data Fig. 6). We find that the pinning of the $4a_0$ unidirectional charge order is weak as it can be spatially manipulated by the tip-induced electric filed. As the STM tip scans along one lattice direction, the positions of the stripes can be spatially shifted along the same direction. Occasionally, the distance between two neighboring stripes changes from $4a_0$ to $5a_0$ (Fig. S15). The properties of the $2a_0\times2a_0$ CDW and the $4a_0$ unidirectional charge order in the normal state are in good agreement with the recent STM work over a wide range of temperatures[14]. Our DFT calculations find that the three-dimensional 2×2×2 CDW arises from phonon softening and electron-phonon coupling in all $AV_3Sb_5$[57]. The phonon spectrum showing mode softening in $CsV_3Sb_5$ is reproduced in Fig. S16, together with a comparison of the large energy scale d$I$/d$V$ spectrum and the DFT band structure. Combining the theory and experiments, we believe the 2×2 CDW order is robust and extends from the SC ground state all the way up to $T_{\mathrm{CDW}} \sim 94$ K[14]. However, we find that the electron-phonon coupling mediated superconductivity alone, obtained by solving the McMillan equation in the reconstructed lattice structure, cannot describe the observed strong-coupling superconductor[57].

The $4a_0$ unidirectional charge order also does not emerge in the electron-phonon coupling picture alone and is most likely also driven by electron correlations, given that it onsets at a substantially lower temperature (~50 K)[14] below the $2a_0\times2a_0$ structural transition at $T_{\mathrm{CDW}}$. While X-ray scattering experiments have observed the bulk 2×2×2 CDW order[58,59], the $4a_0$ charge order has not been detected within the current resolution, leaving the issue of whether or in what form the rotation symmetry breaking $4a_0$ axial CDW exists in the bulk. Since STM/S cannot answer this question, we performed angular-dependent magnetoresistance measurements and clearly observed the twofold symmetry with the anisotropy axis along one of the lattice directions (Extended Data Fig. 7). Surprisingly, the onset of the resistivity anisotropy below ~50 K matches well with the onset temperature $T_{\mathrm{stripe}}$ ~50 K of the $4a_0$ charge order detected by STM[14]. This finding suggests that an incipient rotational symmetry breaking bulk electronic state below $T_{\mathrm{stripe}}$, which can either be a quasi-3D form of the $4a_0$ stripes with interlayer coupling, or a different state that manifests as the $4a_0$ stripes on the Sb terminated surface of the kagome metal $CsV_3Sb_5$.

## Method References

**Acknowledgements**

We thank Ilija Zeljkovic, Stephen Wilson, Jiaxin Yin, and Zhong-Xian Zhao for helpful discussions. The work is supported by grants from the National Natural Science Foundation of China (61888102, 52022105, 11974422 and 11974394), the National Key Research and Development Projects of China (2016YFA0202300, 2017YFA0206303, 2018YFA0305800 and 2019YFA0308500), and the Chinese Academy of Sciences (XDB28000000, XDB30000000, 112111KYSB20160061). Z.W. is supported by the US DOE, Basic Energy Sciences Grant No. DE-FG02-99ER45747.

**Author Contributions:** H.-J.G. designed the experiments. H.C., B.H., Y.X., G.Q, Z.H., Y.Y., C.S., G.L. performed STM experiments with guidance of H.-J.G. H.Y. and H.L. prepared samples. X.D, J.Y, H.Y., S.M., H.Z. and S.N. performed the transport experiments. Z.W., S.Z., H.T. and B.Y. carried out theoretical work. All of the authors participated in analyzing experimental data, plotting figures, and writing the manuscript. H.-J. G. and Z.W. supervised the project.

**Competing Interests:** The authors declare that they have no competing interests.

**Data availability**

Data measured or analyzed during this study are available from the corrsponding author on reasonable request.

## Extended Data Figures

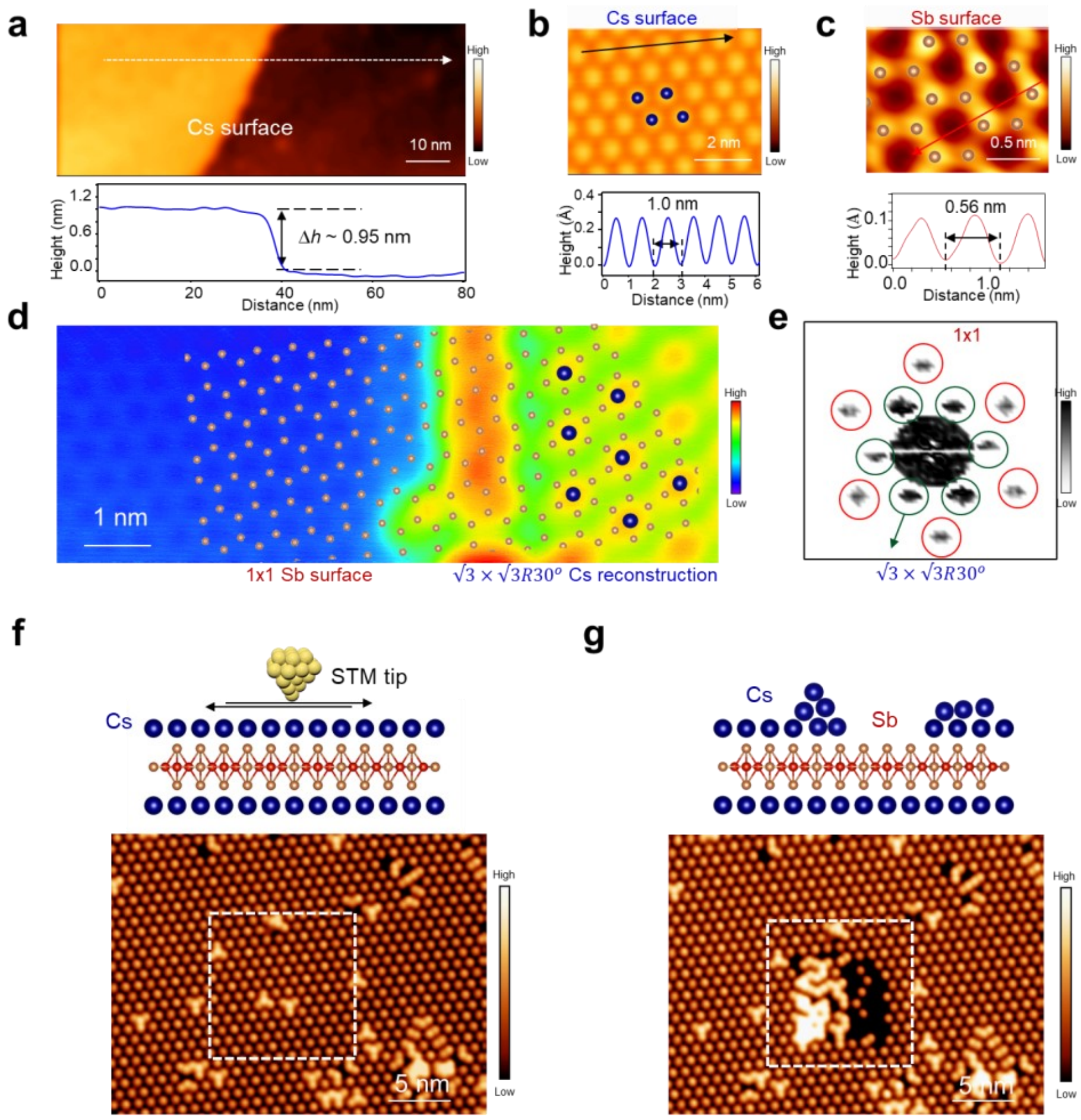

**Extended data Fig. 1 Detailed STM characterization of the Sb and Cs surfaces. a,** Top panel: a typical STM image showing a step edge of Cs surface. Bottom panel: line profile along the white dotted arrow in **a**, indicating that the height of the step edge is ~0.95 nm, which is consistent with the calculated interlayer distance ($V_s$=-1 V, $I_t$=0.1 nA). **b**, Atomically-resolved STM image of Cs surface, showing a hexagonal lattice with a period of 1.0 nm, which is √3 times larger than the lattice constant ($a = b = 0.55$ nm, see Fig. S1a). ($V_s$=-500 mV, $I_t$ =0.5 nA). **c**, Atomically-resolved STM image of Sb surface, showing a honeycomb lattice. The periodicity of the honeycomb lattice is about 0.56 nm, which agrees with the bulk lattice constant ($a = b = 0.55$ nm, see Fig. S1**a**). ($V_s$ =-500 mV, $I_t$ =0.5 nA). **d**, Atomically-resolved STM image of an interface between the top Cs and bottom Sb surfaces (same as in Fig. **1d**). The atomic model is overlaid on the image, showing that each Cs atom sits on top of the Sb honeycomb center ($V_s$=-500 mV, $I_t$ =0.5 nA). **e**, FFT of **d** showing the Cs √3×√3$R$30° reconstruction relative to the Sb 1×1 lattice. **f, g** Top panels: schematics showing STM manipulations to expose the bottom Sb surface. Bottom panels: STM images of Cs surface before (**f**) and after (**g**) STM manipulation, respectively, showing the freshly-obtained bottom Sb surface highlighted by the white dotted square ($V_s$=-500 mV, $I_t$ =0.5 nA).

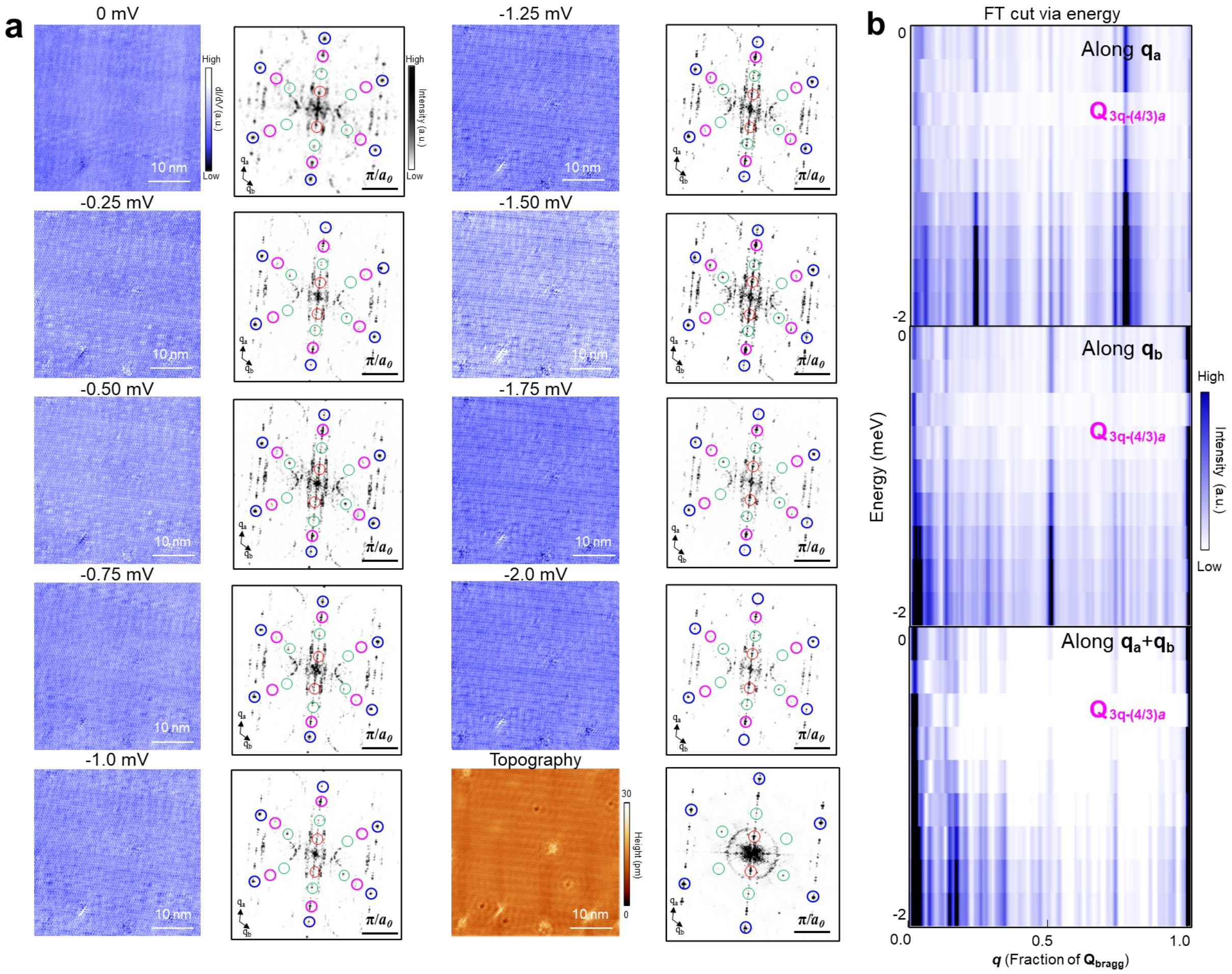

**Extended data Fig. 2. STM topography and d*I*/d*V* maps over a 40 nm × 40 nm region at 300 mK.** **a**, Topography, d*I*/d*V* maps and the intensity of the drift-corrected Fourier transforms at the sample bias from -2 mV to 0 mV, respectively. Each map consists of 500 pixels × 500 pixels. **b**, Energy dependence of the Fourier line cuts along the three directions of the hexagonal zone. ($V_s$=-5 mV, $I_t$=2 nA, $V_{mod}$=0.5 mV).

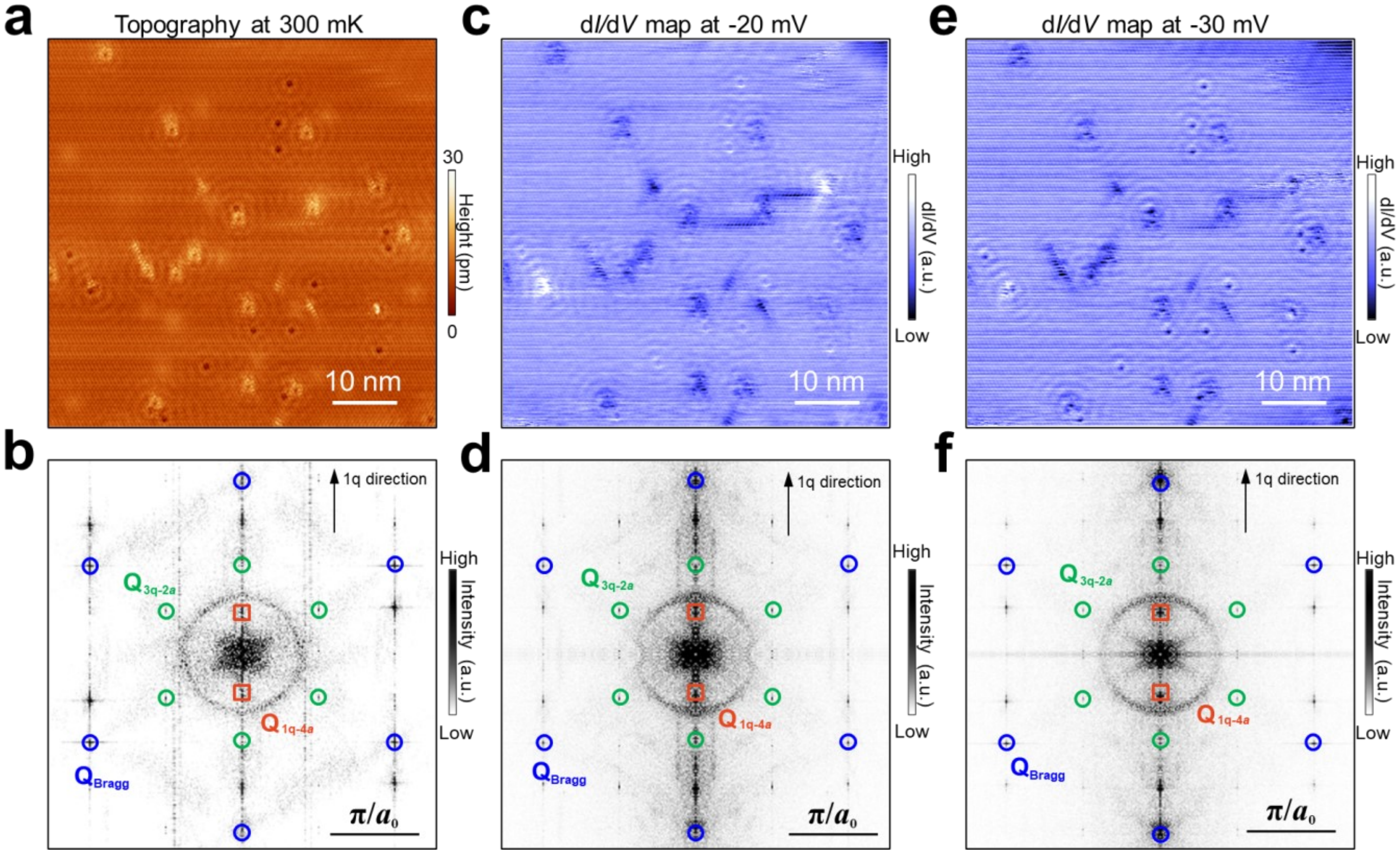

**Extended data Fig. 3. Absence of $4a_0/3$ in high energy d*I*/d*V* maps at 300 mK. a**, Large-scale STM image (60 nm × 60 nm) of the Sb surface obtained at the temperature below $T_c$ (300 mK), where a unidirectional charge order is visible ($V_s$=-20 mV, $I_t$=2 nA). **b**, The magnitude of drift-corrected Fourier transform of **a**, showing clearly the $\mathbf{Q}_{\mathbf{3q\text{-}2}a}$ CDW and $\mathbf{Q}_{\mathbf{1q\text{-}4}a}$ stripe CDW peaks. **c, d** d*I*/d*V* mapping (1024 pixels × 1024 pixels) over the same region at -20 mV and the corresponding magnitude of drift-corrected Fourier transform ($V_s$=-20 mV, $I_t$=2 nA, $V_{mod}$=0.2 mV). **d, f** d*I*/d*V* mapping (1024 pixels × 1024 pixels) over the same region at -30 mV and the corresponding magnitude of drift-corrected Fourier transform ($V_s$=-30 mV, $I_t$=2 nA, $V_{mod}$=0.2 mV).

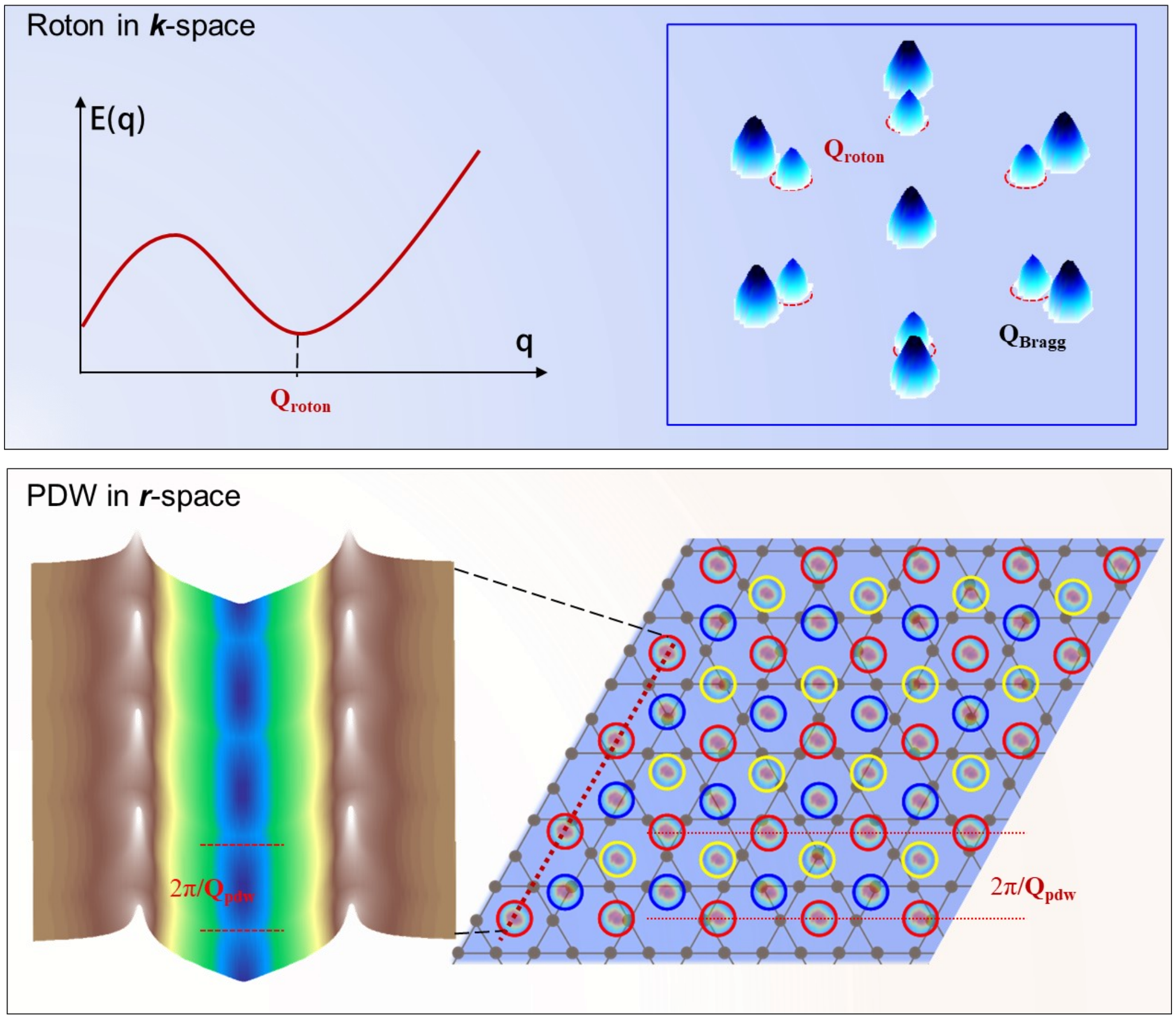

**Extended data Fig. 4. Schematic illustration of the roton-PDW**. Top panel: the roton dispersion and roton minimum at $\mathbf{Q}_{\mathbf{roton}}$= $\mathbf{Q}_{\mathbf{3q\text{-}4}a\mathbf{/3}}$ in the reciprocal lattice. Bottom panel: the 3Q roton-PDW at $\mathbf{Q}_{\mathbf{pdw}}$ = $\mathbf{Q}_{\mathbf{roton}}$ forming a commensurate vortex-antivortex lattice (red, blue and yellow circles) that spatially modulates the tunneling conductance spectra along a line cut.

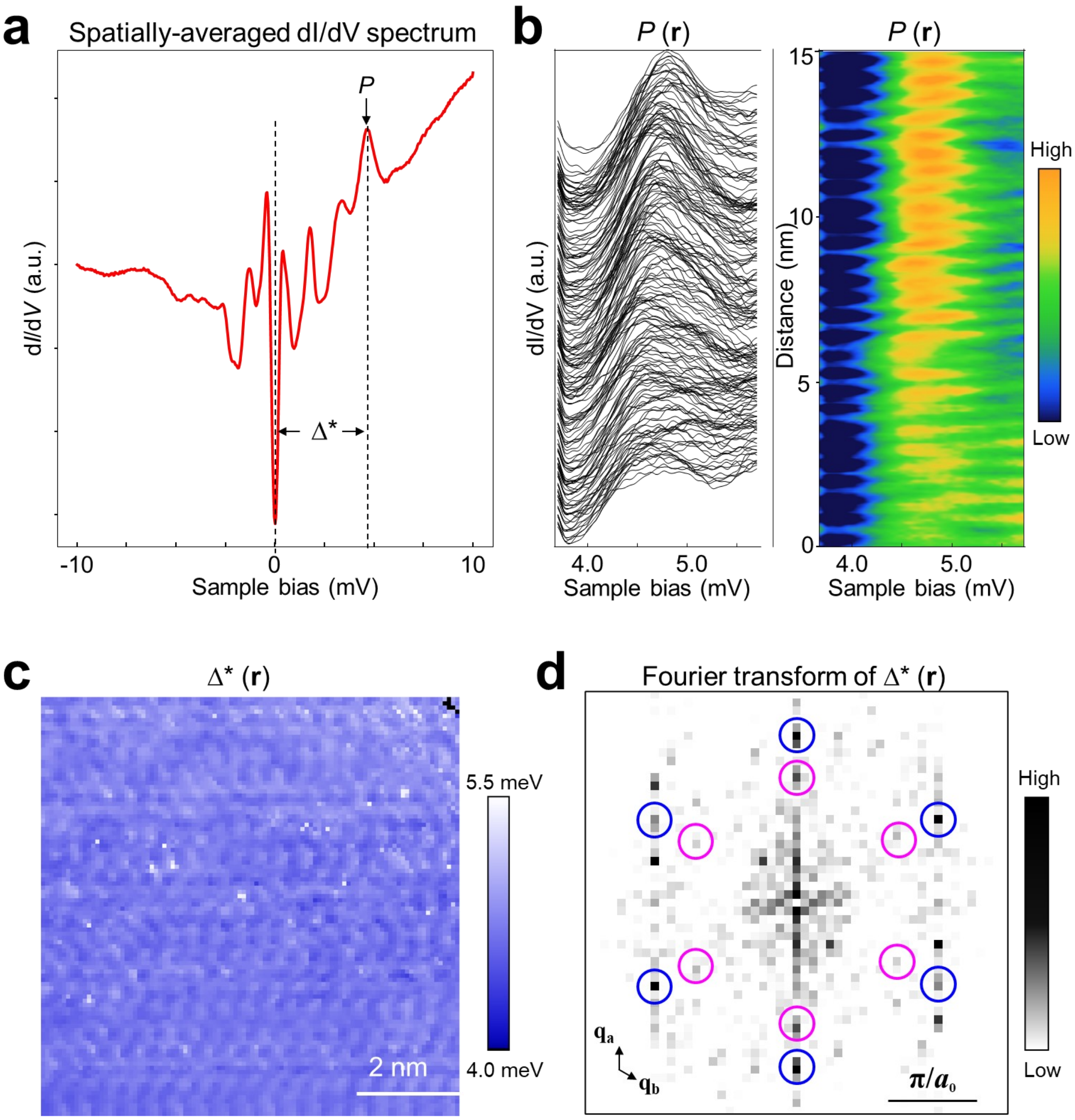

**Extended data Fig. 5. Spatial map of pseudogap and $Q_{3q\text{-}4a/3}$ modulations. a**, Spatially-averaged dI/dV spectrum obtained below $T_c$, exhibiting several peaks in the energy range between 1 mV and 6 mV ($V_s$=-10 mV, $I_t$=1 nA, $V_{mod}$=0.05 mV). The PDW pseudogap peak located near 5 mV is labelled as *P*. **b**, Waterfall and color plot of a d*I*/d*V* line cut, showing spatial modulations of the peak *P* ($V_s$=-3.7 mV, $I_t$=1 nA, $V_{mod}$=0.05 mV). **c**, Spatial gap map of Δ*(**r**), showing the spatial modulations of the pseudogap ($V_s$=-3.7 mV, $I_t$=1 nA, $V_{mod}$=0.05 mV). **d**, Fourier transform of the pseudogap map showing peaks at the PDW vectors $\mathbf{Q}_{3q\text{-}4a/3}$ circled in magenta.

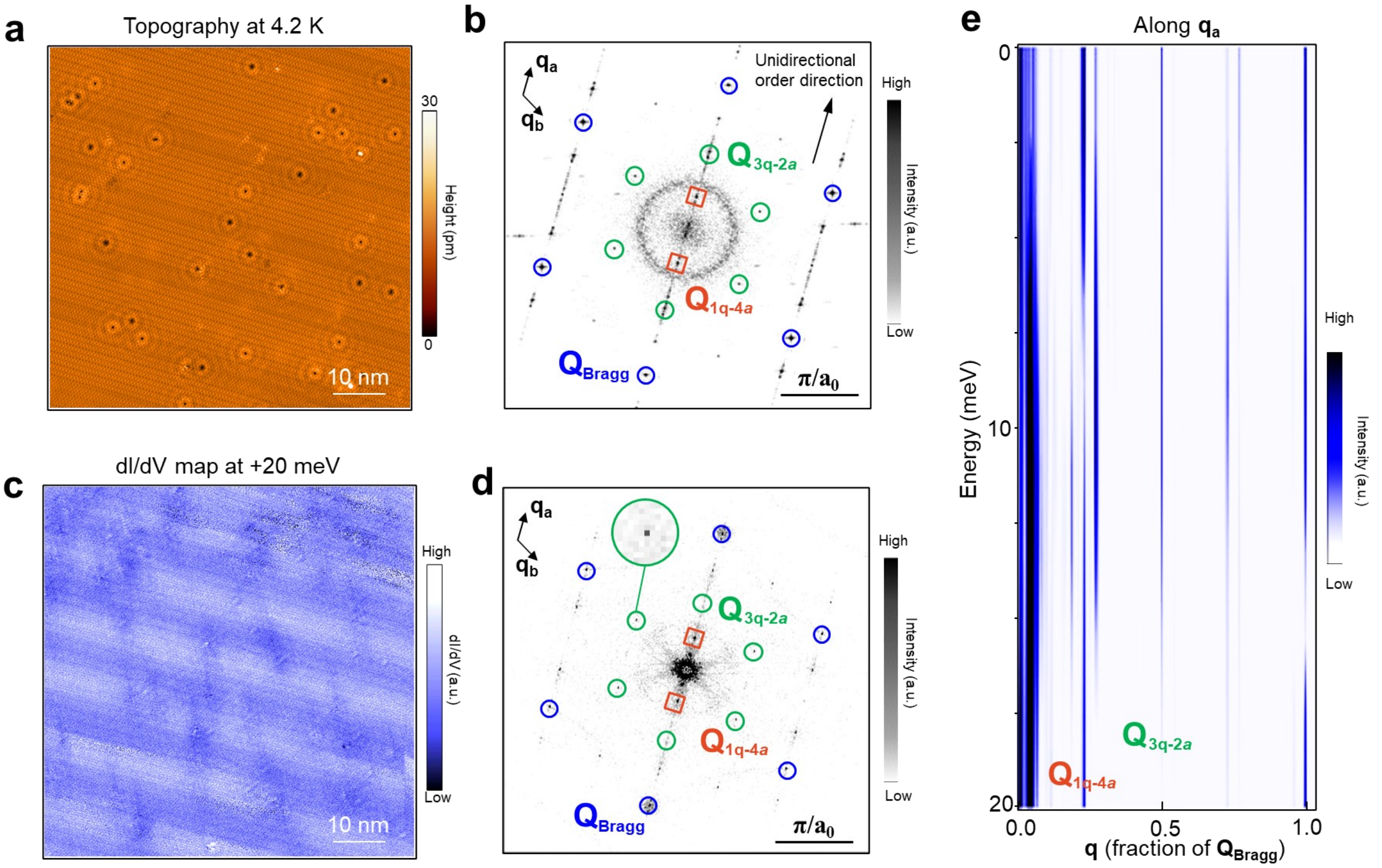

**Extended data Fig. 6. Charge ordered normal state in $CsV_3Sb_5$ above $T_c$.** **a**,**b** Large-scale STM topography of Sb surface obtained at 4.2 K and the magnitude of drift-corrected Fourier transform, showing $2a_0 \times 2a_0$ and $4a_0$ striped CDW peaks at wave vectors $\mathbf{Q}_{3q\text{-}2a}$ and $\mathbf{Q}_{1q\text{-}4a}$ ($V_s$=-90 mV, $I_t$=2 nA). **c,d** d$I$/d$V$ mapping of **a** at 20 mV and the magnitude of drift-corrected Fourier transform, respectively ($V_s$=-90 mV, $I_t$=2 nA, $V_{mod}$=0.5 mV). **e.** Energy dependence of the Fourier line cuts along $\mathbf{q_a}$ directions, showing that peaks at $\mathbf{Q}_{3q\text{-}2a}$ and $\mathbf{Q}_{1q\text{-}4a}$ at 4 K are non-dispersive ($V_s$=-90 mV, $I_t$=2 nA, $V_{mod}$=0.5 mV).

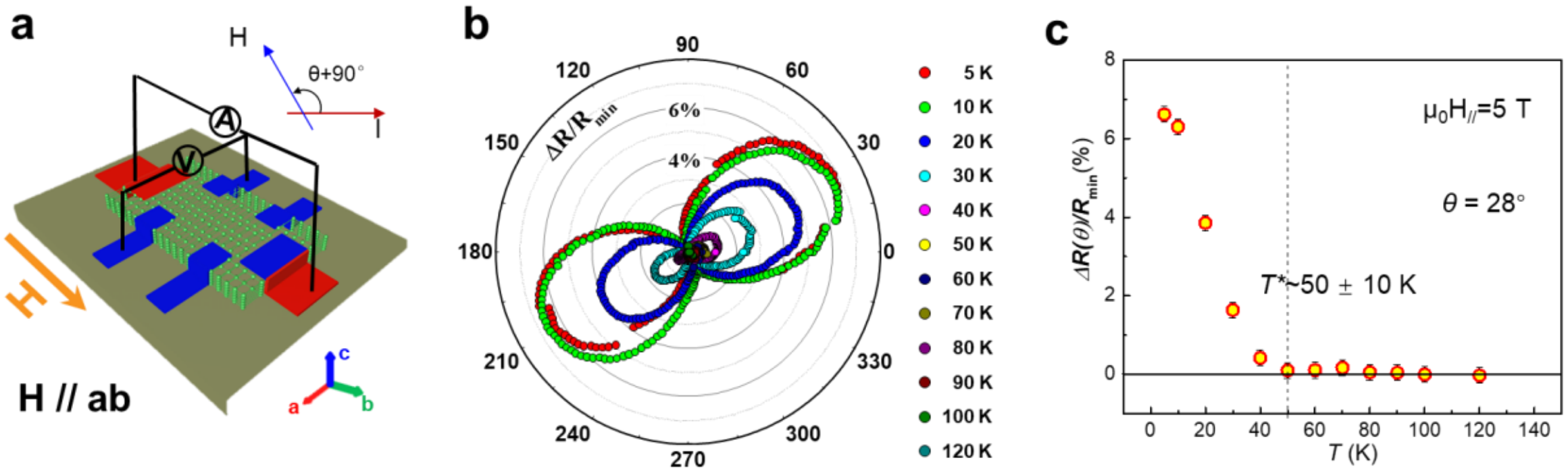

**Extended Data Fig. 7. Normal state angular-dependent magnetoresistance. a**, Schematic of the in-plane resistance measurement under a 5 T magnetic field by rotating the sample along c axis of the single crystal. **b,** Angular plot of the normalized anisotropic magnetoresistance ($\Delta R/ R_{min}, \Delta R = R(\theta) - R_{min}$), showing two-fold symmetry at the temperature below ~50 K. $\theta$ is defined in **a**. **c,** Temperature dependence of the angular-dependent of $\Delta R/ R_{min}$ at the angle of 28°, showing the onset of twofold rotational symmetry below $T$*~50 ± 10 K.